%% file: hcp06-rare-decays.tex
\documentclass[final,numberedheadings,sort&compress]{aipproc}

\layoutstyle{8x11single}

\graphicspath{{./graphics/}}

\include{definitions}

\usepackage{graphicx}
\usepackage{dcolumn}
\usepackage{bm}
\usepackage{setspace}

\usepackage{amssymb,amsmath}

\begin{document}
\title{Rare Decays of Heavy Flavor at the Tevatron}


\author{V.~Krutelyov\\ (On behalf of CDF and D\O\ Collaborations)}{
  address={Department of Physics, UCSB, Santa Barbara, CA 93105\\slava77@fnal.gov}
}

\begin{abstract}
In this report I review recent results in the field of rare decays 
at the Tevatron \cdf\ and D\O\ experiments.
The presentation is focused on  rare decays of charm and bottom mesons
with two muons in the final state.
This includes improvements over the previously available limits on the 
following branching ratios:
$\mathcal{B}(\dpmmp)< 4.7\EE{-6}$, 
$\mathcal{B}(\bsmmphi)< 3.2\EE{-6}$, 
$\mathcal{B}(\bsmm)< 1 \EE{-7}$, and $\mathcal{B}(\bdmm)< 3 \EE{-8}$ 
all at the $90\%$ confidence level.
Also reported are the first direct observation
of $D_s^+\to\phi\pi^+\to\mm\pi^+$ with a significance above background of over 7 standard 
deviations 
and evidence of $D^+\to\phi\pi^+\to\mm\pi^+$ with a significance of 3.1 and 
$\mathcal{B}(D^+\to\phi\pi^+\to\mm\pi^+)=(1.75\pm0.7\pm0.5)\EE{-6}$.

\end{abstract}

\maketitle


\section{Introduction}
Historically, rare decays of hadrons have been one of the main interests in the 
program of high energy physics (further information on this
subject can be found in Refs.~\cite{PDBook,HFAG:2006bi,Burdman:2001tf,Krutelyov:2005tc}).
Decays with substantially low rates (typically with branching ratios below $10^{-5}$),
rare decays, not only correspond to properties and conservation laws of
the Standard Model (SM), but also provide an invaluable tool in defining
or limiting models of the New Physics (NP).
Owing to the mass hierarchy of quarks and specifics of the quark mixing in the SM,
described by the Cabibbo-Kobayashi-Maskawa (CKM) matrix,
rare decays of heavy flavor hadrons are 
of the most interest.

Based on  a final state  rare decays 
can be classified as
charmless hadronic, radiative, and leptonic.
Here I focus on leptonic decays with two muons in the final state,
corresponding to
Flavor Changing Neutral Currents (FCNCs).

The FCNCs are suppressed and appear only at the loop level 
in the SM and vanish in the limit of 
zero quark masses (Glashow-Iliopoulos-Maiani or GIM mechanism).
Suppression of the FCNCs is a paramount property of any viable
NP model.
Specific to flavor, the FCNCs correspond
to $b\to s (d)$ and to $c\to u$ transitions for the bottom and charm decays
respectively.
While excellent agreement of the SM and experiment 
in processes like $b\to s\gamma$, $b\to s \lplm$,
and $K\to \pi\nu\overline{\nu}$ already constrains many NP models,
further knowledge on FCNC decays is needed in search for the NP.

Having a high experimental sensitivity to rare decays of heavy flavor mesons
requires
ability to produce a large number of such mesons as well as 
to effectively discriminate the signal from backgrounds.
Bottom and charm mesons can be produced
in abundance both at the \ee\ $B$- or charm factories
and at the hadron colliders, like the Tevatron.
Although the production rates
of the heavy flavor are higher at the Tevatron
than at the \ee\ flavor factories 
by far more abundant light flavor production 
puts the Tevatron on a similar level of sensitivity.
Ability to produce all flavors
puts the Tevatron on the forefront in rare decays of hadrons
not easily accessible at $B$-factories, like \bs\ or $\Lambda_b$.

Recent results reported by the CDF and D\O\ are presented
below. 
I begin with highlights of detector features
and common analysis methods.
I then overview particular searches:
for \dmm\ at \cdf\  and for \dpmmp\
at D\O;
for \bsmmphi\ at D\O, 
and for \bmm\ at both experiments.
This is followed by a brief conclusion.

\section{Detectors and common approach to rare decay search}
\label{sec:det}
The collider detectors a the Tevatron, \cdf\ and D\O, described
elsewhere~\cite{Abazov:2005pn,Acosta:2004yw}, are general purpose detectors each having parts
with similar functions.
Relevant to the analyses presented here are the tracker immersed in
the solenoidal field used for precision momentum measurement of
charged particles; the silicon vertex detector used to
effectively select displaced vertices characteristic to decays
of heavy flavor; 
and the muon detectors (located behind  calorimeters and additional steel absorbers) 
used to identify muons.
Rapidity and momentum coverage for muons are 
$|\eta|\lesssim 2 (1)$ and $p_T\gtrsim 3(1.5)~\GeVc$ for D\O\ (\cdf).
Better rapidity acceptance
and higher quality of the muon identification in D\O\
is levered by better tracking precision
 and lower momentum thresholds
for muons at \cdf, which allows both experiments to perform
at a similar level in searches for exclusive decays like $H_{b(c)}\to\mm+X$.

High purity of dimuon selection allows both experiments to effectively 
trigger on dimuon events, which constitute the dimuon samples
used to analyze the rare decays.
In addition to the dimuon triggers, the CDF employs the displaced (two-)
track trigger which allows to select heavy flavor decays based on the
tracks of the decays products alone.

The analyses presented here employ a
relative normalization, where an abundant mode ($H_y\to Y$)
is used to estimate the rate of a rare
mode ($H_x\to X$) collected in the
same sample. 
The branching ratio of the rare mode is given by
$$
\mathcal{B}(H_x\to X) = \mathcal{B}(H_y\to Y) 
\frac{N_{X}}{N_Y} \frac{\epsilon_Y}{\epsilon_X} \frac{f_y}{f_x},
$$
where $N_{X(Y)}$ is the number of  events or an upper limit in the rare
(normalization) mode; 
$\epsilon_{X}$ is the total efficiency,
a fraction of  observed $H_x\to X$ events
relative to all such events produced
(same for $\epsilon_{Y}$);
$f_{x (y)}$ is the relative production fraction of $H_{x(y)}$.
A benefit of this approach is that if kinematics of both modes is similar,
a large part of the systematic uncertainty cancels in the ratio.

The strategy of the analyses is the following.
First,
pick dimuon events passing baseline selections, like 
a noticeable momentum to expect a displaced decay vertex
and a set of quality requirements to muons
and the decay vertex.
This gives a sample with an order of $10^3 - 10^4$ events completely dominated by background.\footnote{
The potential sources of combinatorial
background are
sequential semileptonic $b\to c\to s$ decays, double semileptonic
$bb\to \mm X$ decays, and events with charged particles
mis-identified as muons (fake muons).}
At this point the events 
with mass near the signal meson mass 
are hidden and the optimal choice of cuts
is based on events in the sidebands used
to predict backgrounds in the signal region
and the signal itself is modeled using the Monte Carlo (MC) simulation.
Once the optimization is done events
in the signal window are uncovered
and the limit on the branching ratio is set.

\section{Rare decays of charm mesons}
\label{sec:cRare}
The $c$-FCNCs in the SM correspond to 
the loop diagrams
with $\{d, s, b\}$ in the loop
where the contribution from non-vanishing
$s$-quark mass dominates.
The GIM suppression works in this case due to 
small masses of $d$- and $s$- quarks and a CKM suppression
of $b$-quark contribution.
As a result the rare decays are  dominated by
long distance interactions ($\phi$ or other internal resonance
decays to \mm)~\cite{Burdman:2001tf}.
The $c$-FCNCs can be enhanced in the NP models to rates
as high as the present experimental sensitivity.
A search for \dmm\ at \cdf~\cite{Acosta:2003ag}
and a search for \dpmmp\ recently reported by the D\O~\cite{D0n5038:dpmmp} are
reviewed briefly below.

\subsection{Search for \dmm\ at \cdf}
\label{sec:dmmCDF}
The branching ratio of the  \dmm\ in the SM is about $10^{-13}$ 
(the short distance contribution is only about $10^{-19}$)~\cite{Acosta:2003ag,Burdman:2001tf},
which in addition to the GIM suppression also has a helicity
suppression factor of $(m_{\mu}/m_{D})^2$.
This value is substantially lower than
experimental limits of about $10^{-6}$ (including the most recent 
measurements~\cite{Aubert:2004bs}).

A search for \dmm\ decay was performed at \cdf\ using 68~\pbin~\cite{Acosta:2003ag}
with normalization to \dpp\ ($\mathcal{B}\sim 1.4\EE{-3}$) with events
from the  two-track trigger sample.
The data with reconstructed \mm\ in the region within $ 22~\MeVcc$  ($2\sigma$)
of the mass of $D^0$
were hidden
during the optimization.
Two backgrounds contributed in this case:
combinatorial (estimated from the high-mass sideband)
and mis-identification of \dpp\ as \dmm.

To discriminate signal from background the following variables were used:
azimuthal angle between the extrapolated positions of the 
tracks at muon chambers ($\Delta\phi$), impact parameter of the candidate ($d_{xy}$) 
and its decay length projected onto its transverse momentum ($L_{xy}$)
both in the plane transverse to the beam.
Values of the cuts  where chosen to maximize 
$S/(1.5+\sqrt{\nbgd})$, where $S(\nbgd)$
is the number of signal (background) events corresponding
to the best limit at $99.7\%$ confidence level (\CL).
The optimization yields\footnote{Note that at the trigger
level the candidates are required to have $L_{xy}>200~\um$.}: 
$|\Delta\phi|>0.085$, $|d_{xy}|<150~\um$,
and $L_{xy}< 0.45~\cm$.

After applying the optimal requirements 5 events are left in the high-mass
dimuon sideband and $1412\pm 54$ events are observed in \dpp\ mode,
which corresponds to expected \nbgd\  
of $(1.6\pm 0.7)_{\mathrm{comb.}} + (0.22\pm 0.02)_{\mathrm{mis-id.}} = 1.8\pm 0.7$.
No events were observed in the data, corresponding to the 
upper limit of $2.5\EE{-6}$ at $90\%~\CL$
This measurement was superseded by a limit of 
$1.3\EE{-6}$ at $90\%$~\CL\ reported in Ref.~\cite{Aubert:2004bs}.
With more than $1~\fbin$ of data collected by \cdf\
a substantial improvements to the limit is expected with an updated analysis.

\subsection{Search for \dpmmp\ at D\O}
\label{sec:dpmmpD0}
Decays of $D^+$ and $D^+_{s}$  to $\mm h^+$,
where $h$ is a kaon or a pion, are driven by long distance interactions in the SM.
Their branching ratios in the SM range from $6.1\EE{-6}$ for
\dsmmp, and $1\EE{-6}$ for \dpmmp\ to $7.1\EE{-9}$ for \dpmmk, 
with significant
short-distance FCNCs only in \dpmmp\ ($\mathcal{B}\sim 9.4\EE{-9}$) 
and \dsmmk\ ($\mathcal{B}\sim 9\EE{-10}$)~\cite{D0n5038:dpmmp}.
Previous searches for these modes
reveal limits of the order of  $10^{-5}$.

A search for short-distance (off $\phi$-resonance) decay \dpmmp\
was done by D\O\ using 1~\fbin\ of data~\cite{D0n5038:dpmmp}.
The analysis was performed in two stages: the branching ratio of \dpmmp\ on $\phi$
resonance (with $0.96<\Mmm<1.06~\GeVcc$) 
was measured first normalized to \dsmmp;
then the \dpmmp\ off $\phi$ resonance was searched for normalized to
the resonant part.
The data sample was collected using the dimuon trigger.

The following variables 
were used to discriminate 
signal from background.
Decay vertex significance, $S_D$, and significance of the impact
parameter of the pion track, $S_{\pi}$, both in the plane transverse to the beam.
Collinearity or angle between the $D$-candidate momentum and
the vertex displacement from the beamline, $\Theta_{D}$.
Track isolation of the candidate, $I_D=p(D)/\sum{p}$, where the sum
is over tracks with $\Delta R\equiv \sqrt{\Delta\eta^2 + \Delta\phi^2}<1$ 
relative to the candidate momentum.
Quality of the candidate, 
$\mathcal{M} = \chi^2_{\mathrm{vtx}} + 1/p^2_{T,\pi} + \Delta R^2_{\pi}$,
where $\chi^2_{\mathrm{vtx}}$ is for the fit to the decay vertex,
$p_{T,\pi}$ is a transverse momentum of the pion in~\GeVc,
and $\Delta R_{\pi}$ is the distance of the pion
track from the dimuon system.

The optimization of these cuts was performed to maximize
$\epsilon/(0.82+\sqrt{\nbgd})$, where $\epsilon$ is the signal efficiency
(relative to preselections of $I_D>0.4$, $S_D>3$, $S_\pi>0.5$, and $\Theta_D<50$~mrad),
which corresponds to the maximum expected upper limit at $90\%$~\CL.
The cuts chosen as a result of the optimization
are $I_D>0.44(0.71)$, $S_D>3.4(9.4)$, $S_\pi>0.57(1.8)$, $\Theta_D<32(7)$~mrad, 
and $\mathcal{M}<6.1(2.6)$
for the on-resonance (off-resonance) measurement.

\begin{figure}
\includegraphics[width=0.45\textwidth]{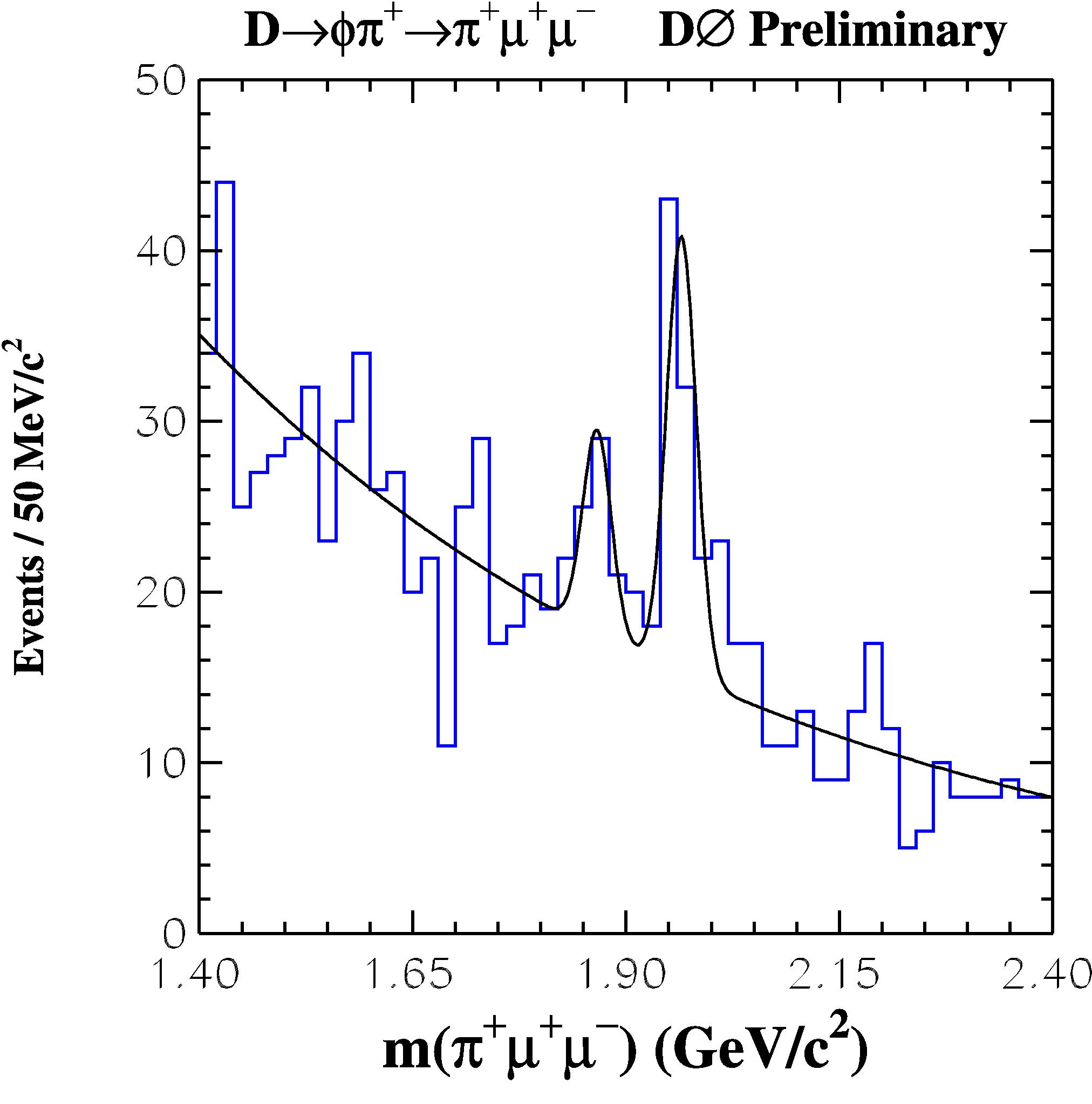}
\includegraphics[width=0.45\textwidth]{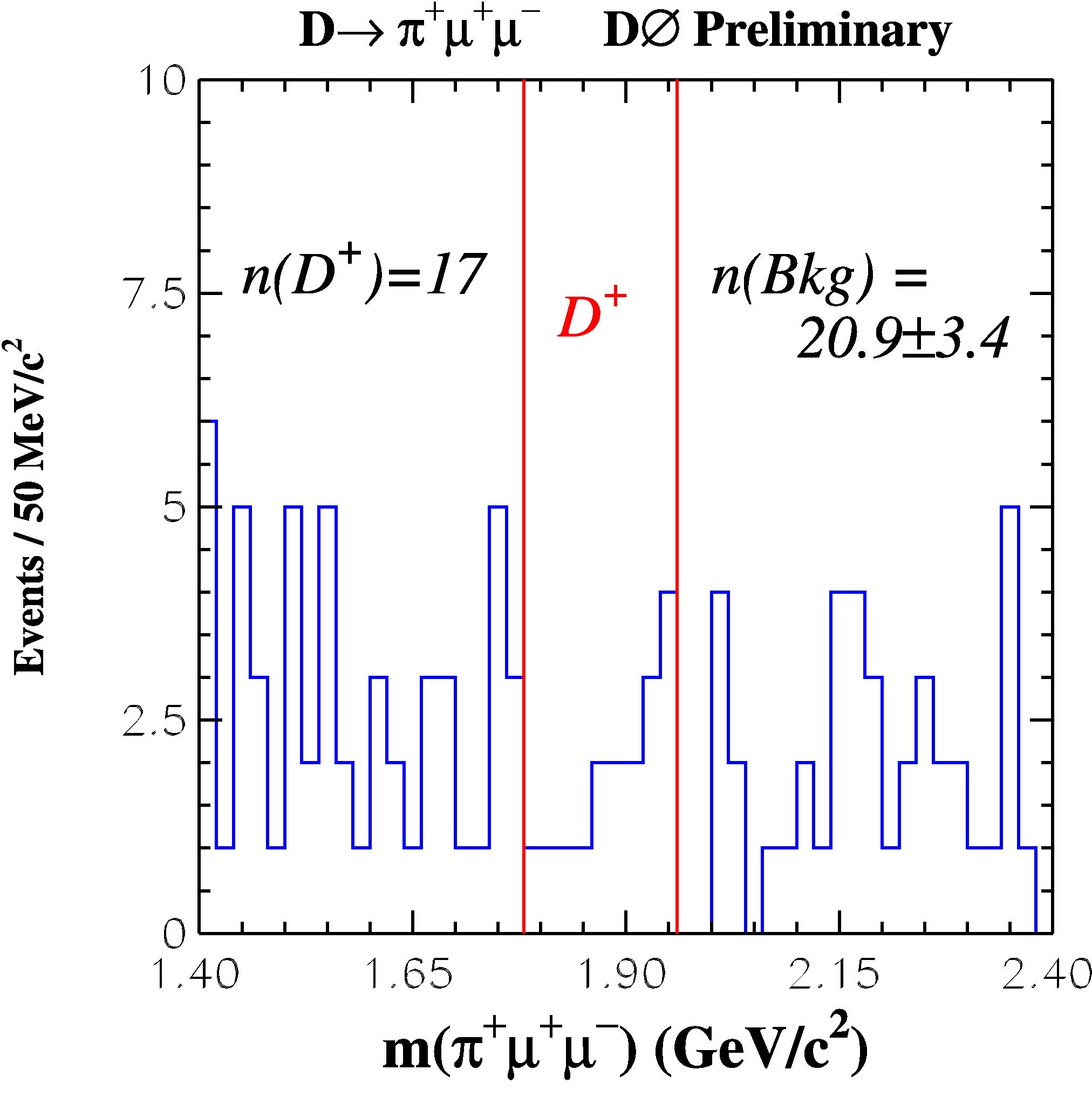}
\caption{\label{fig:d0dmm} The $m_{\pi^{+}\mm}$ mass spectrum for the optimized
on-resonance (left) and off-resonance (right) selection criteria.
The curve on the left plot is the result of the fit to contributions from $D^+$,
$D^{+}_{s}$, and combinatoric backgrounds.}
\end{figure}

The $D$-candidate mass spectra for on- and off-resonance
after the optimal selections are shown in Fig.~\ref{fig:d0dmm}.
For the on-resonance case the fit yields
$65\pm 11$ $D^+_s$ events and $26\pm 9$ $D^+$ events, which corresponds
to about $7\sigma$  and $3.1\sigma$ significance for $D_s^+$ and $D^+$ respectively.
The corresponding measurement is
$\mathcal{B}(D^+\to\phi\pi^+\to\mm\pi^+) = (1.75\pm 0.7\pm 0.5)\EE{-6}$,
consistent with simple factorization of two sequential decays,
and can be compared to the recent CLEO-c measurement
of $(2.7\ase{3.6}{1.8}\pm 0.2)\EE{-6}$.
Off-resonance, the number of observed events in the signal window
is 17 consistent with  $\nbgd = 20.9\pm3.4$,
corresponding to
$\mathcal{B}(\dpmmp)<4.7\EE{-6}$ at $90\%$~\CL

\section{Rare decays of bottom  mesons}
\label{sec:bRare}
Rates of the $b$-FCNCs in the SM
are higher than of the similar $c$-FCNCs and 
some are within the present experimental sensitivity, like $b\to s\gamma$ and
$b\to s\lplm$ in $B^+$ and $B^0$ decays~\cite{PDBook}.
This is because the GIM suppression is not effective due to the top-quark in the loop.
In the NP models the $b$-FCNCs can be enhanced by large factors.
The results of searches for \bsmmphi\ at D\O\ and for \bmm\ at both experiments are discussed 
below.

\subsection{Search for \bsmmphi\ at D\O}
\label{sec:bsmmphiD0}
In the SM, the rate of the \bsmmphi, neglecting the interference
effects with  $\bs\to\jp\phi$ 
($\mathcal{B}\sim 6\EE{-5}$~\cite{HFAG:2006bi}) 
and $\bs\to\psi(2S)\phi$ ($\mathcal{B}\sim 4\EE{-6}$~\cite{HFAG:2006bi})
resonant decays, is predicted to be about $1.6\EE{-6}$~\cite{Abazov:2006qm}.
The previously published limit
of $\mathcal{B}(\bs\to\mm\phi)<6.7\EE{-5}$ at $95\%~\CL$  comes from CDF~I.
This decay is related to $b\to s\lplm$ transition
and by its properties is similar to the  observed
$\bd\to K^{\star}\lplm$ ($\mathcal{B} = (1.46\pm 0.25)\EE{-6}$~\cite{HFAG:2006bi}).
Observation of \bsmmphi\ and analysis of its kinematics
will add important information on the flavor dynamics of $b$-FCNCs.

The D\O\ Collaboration has performed a search for \bsmmphi\ 
using $0.45~\fbin$~\cite{Abazov:2006qm} with $\bs\to\jp\phi$
used for normalization.
The events were collected from the dimuon trigger sample in the
range of $0.5< \Mmm < 4.4~\GeVcc$ and with 
the $\phi$-candidate mass within $1008 <m_{\phi}< 1032~\MeVcc$.
With an addition of a good decay vertex and $p_T^{B}> 5~\GeVc$ requirements
this defined the baseline sample of 2602 events.
At this point the events in the signal window within $188~\MeVcc$  ($2.5\sigma$)
from the world-average $m_{\bs}$ were hidden from the optimization.
The $\bs\to\jp\phi$ events are required to pass the same cuts as the signal
except for the dimuon mass to be consistent with \jp.
In the signal mode events with  $2.72< \Mmm < 4.06~\GeVcc$ 
corresponding to charmonium resonances were removed.

The following variables were used to further
discriminate signal from background with values chosen
to maximize $\epsilon/(1+ \sqrt{\nbgd})$ (corresponding to the
best limit at $95\%~\CL$), where $\epsilon$ is the signal efficiency
relative to the baseline selections.
Collinearity or pointing angle between the candidate momentum
and the displacement of the decay vertex from the beamline, 
$\Theta_{B}$.
Significance of the candidate decay length, $S_{B}$.
The track isolation, $I_{B} = p_{B}/(p_{B}+ \sum{p})$, where
the sum is taken over non-candidate tracks momenta with $\Delta R<1$ relative to the direction
of the candidate momentum $p_{B}$.
The optimal cuts were shown to be:
$S_{B}>10.3$, $I_{B}>0.72$, and $\Theta_{B}<0.1~\mrm{rad}$.

In the data
$73\pm 10 \pm 4$ and zero events pass the optimal selections in $\bs\to\jp\phi$ and the signal modes
respectively
with
an expected \nbgd\ of $1.6\pm 0.4$
and $\epsilon \approx 54\%$.
The resulting limit is $\mathcal{B}(\bsmmphi)< 3.2\EE{-6}$ at $90\%~\CL$,
which is only about a factor of two above the SM value and is 
substantially better than the previously published limit.

\subsection{Search for \bmm}
\label{sec:bmm}
In the SM, the branching ratio of \bs\ (\bd) decay to \mm\ 
is about $4\EE{-9}$ ($1\EE{-10}$)~\cite{Krutelyov:2005tc},
which is helicity-suppressed
by a factor of $(m_{\mu}/m_B)^2$.
Compared to the \bsmm, the \bdmm\ mode is further suppressed
by  a factor of $(V_{td}/V_{ts})^2 \sim 0.04$.
The present experimental sensitivity to \bsmm\ 
of about $10^{-7}$
already becomes important in constraining NP scenarios.
In the MSSM with large $\tan{\beta}$ the dominant
contribution is from a heavy neutral Higgs 
exchange (proportional to $\tan^6{\beta}/m_{H}^4$),
which, considering other constraints, can be
as large as the current sensitivity.
Observation of the \bsmm\ at the Tevatron would be an unequivocal
indication of the NP.

The D\O\ reported a search for \bsmm\
using $300~\pbin$ (the first sample)~\cite{Abazov:2004dj},
and a sensitivity study using additional
$400~\pbin$ (the second sample)~\cite{D0n5009:bsmm}.
The CDF reported results
based on $780~\pbin$~\cite{CDF8176:bsmm},
using the method as in the previous
study detailed in~\cite{Krutelyov:2005tc}.
Both experiments used dimuon trigger samples
and normalized to \bpjpk.


\subsubsection{Search for \bsmm\ at D\O}
\label{sec:bmmD0}
With a dimuon mass resolution near \bs\ mass 
of about $90~\MeVcc$  \bs\ and \bd\ can not be separated.
Thus, assuming the contribution from \bd\ is small, the search is targeted on \bs.
The preselection of \bs\ candidates is made from
dimuon events with $4.53 < \Mmm < 6.15~\GeVcc$
and each muon with $p_T>2.5~\GeVc$.
The  candidates are required to have a good decay vertex
and $p_T^B> 5~\GeVc$.
These selections leave about $4\EE{4}$ events in the first sample
(slightly more in the second sample).
At this point events within $270~\MeVcc$ from \bs\ mass are hidden
from the optimization procedure.

The cuts on variables 
$S_B$, $\Theta_{B}$, and $I_B$
are chosen to maximize $\epsilon/(1+ \sqrt{\nbgd})$ 
(denoted  as in Section~\ref{sec:bsmmphiD0}).
The optimal cuts for the first (second) data sample are:
$I_B>0.56(0.59)$, $S_B>18.5(19.5)$, and $\Theta_B < 0.2(0.18)$~rad.

The events in the \bpjpk\ normalization mode are required to pass the same
selections as the signal mode, 
except for the dimuon mass to be consistent with
\jp, and an addition of kaon track with $p_T>0.9~\GeVc$.

After applying the optimal selections
the number of \bpjpk\ events is $741\pm 31 \pm 22$  ($899\pm 37$)
in the first (second) data sample.
In the signal window within $180~\MeVcc$ from
the \bs\ mass the \nbgd\ is $4.3\pm 1.2$ ($2.2\pm 0.7$) in the first (second)
sample.
In the signal region 4 events are observed in the first
data sample, corresponding to  $\mathcal{B}(\bsmm)<4\EE{-7}$ at $95\%~\CL$
The signal region in the second sample remains hidden pending
a decision on improvements to the analysis.
The expected limit using both samples is
$\langle \mathcal{B}(\bsmm)\rangle <2.3\EE{-7}$ at $95\%~\CL$

\subsubsection{Search for \bmm\ at \cdf}
\label{sec:bmmCDF}
The baseline sample
is selected from events with dimuon mass within
$4.669< \Mmm < 5.969~\GeVcc$ and each muon with
$p_T> 2~\GeVc$.
The $B$-candidates are required to have $p_T^B> 4~\GeVc$, 
rapidity in range $|y^B|<1$,
and a decay vertex displaced from the production
vertex by $2\sigma$.
In addition the $B$-candidates are required to have
a pointing angle (between the candidate momentum and the decay vertex displacement) 
in range $\Theta< 0.7$~rad
and a track isolation ($I = p_T^B/(p_T^B + \sum {p_T})$, where the sum is taken
over non-candidate tracks with $\Delta R< 1$ relative to the momentum of the candidate)
in range $I>0.5$.
At this point, out of 23066 events left,
events with $5.169 < \Mmm < 5.469~\GeVcc$
(the signal window is $\pm 60~\MeVcc$ or $2.5\sigma$
around $m_{\bs}$ or $m_{\bd}$)
are hidden from the optimization procedure.

\begin{figure}
\includegraphics[viewport=0 0 550 539,clip,width=0.45\textwidth]{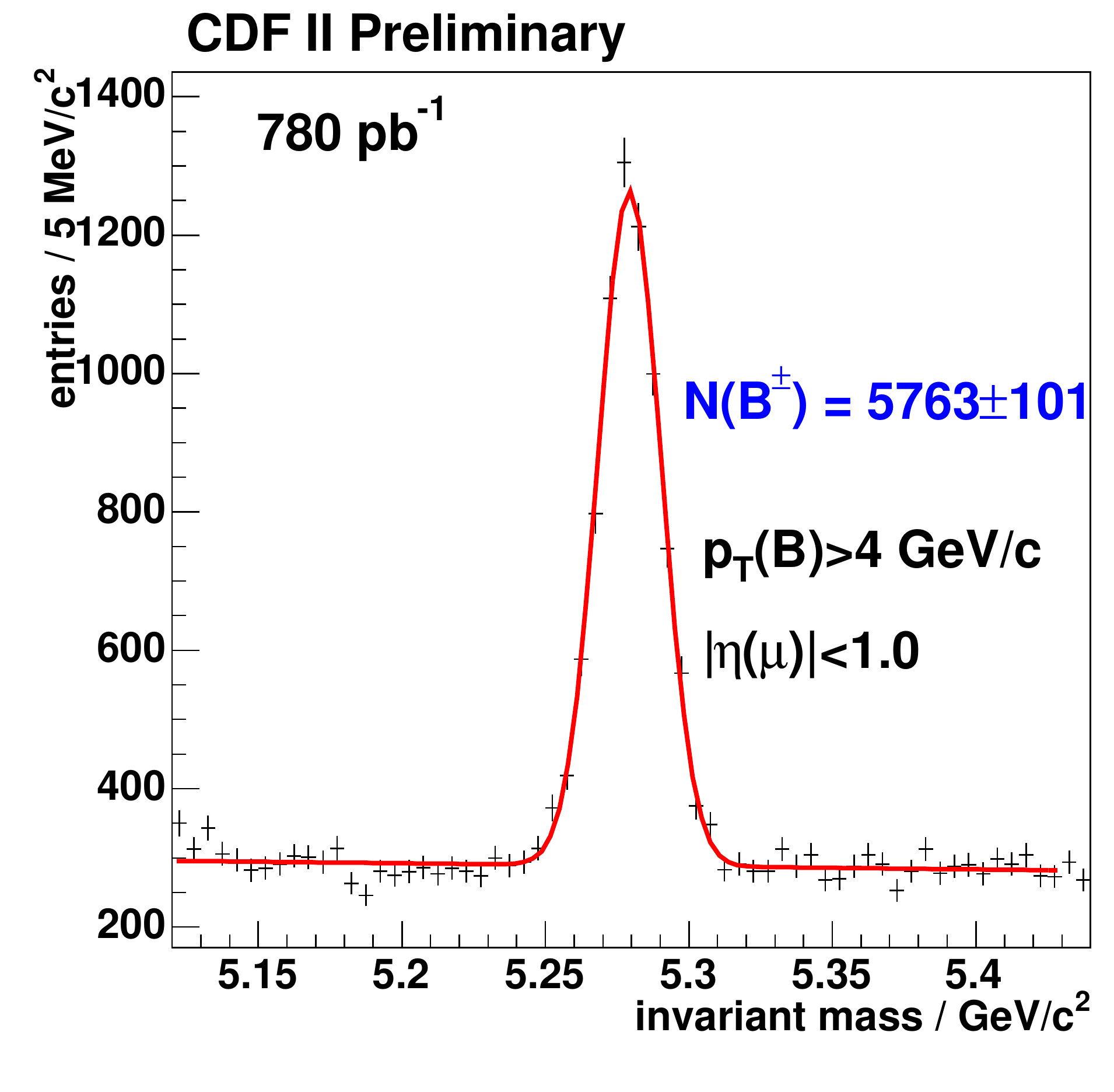}
\includegraphics[viewport=0 0 555 539,clip,width=0.45\textwidth]{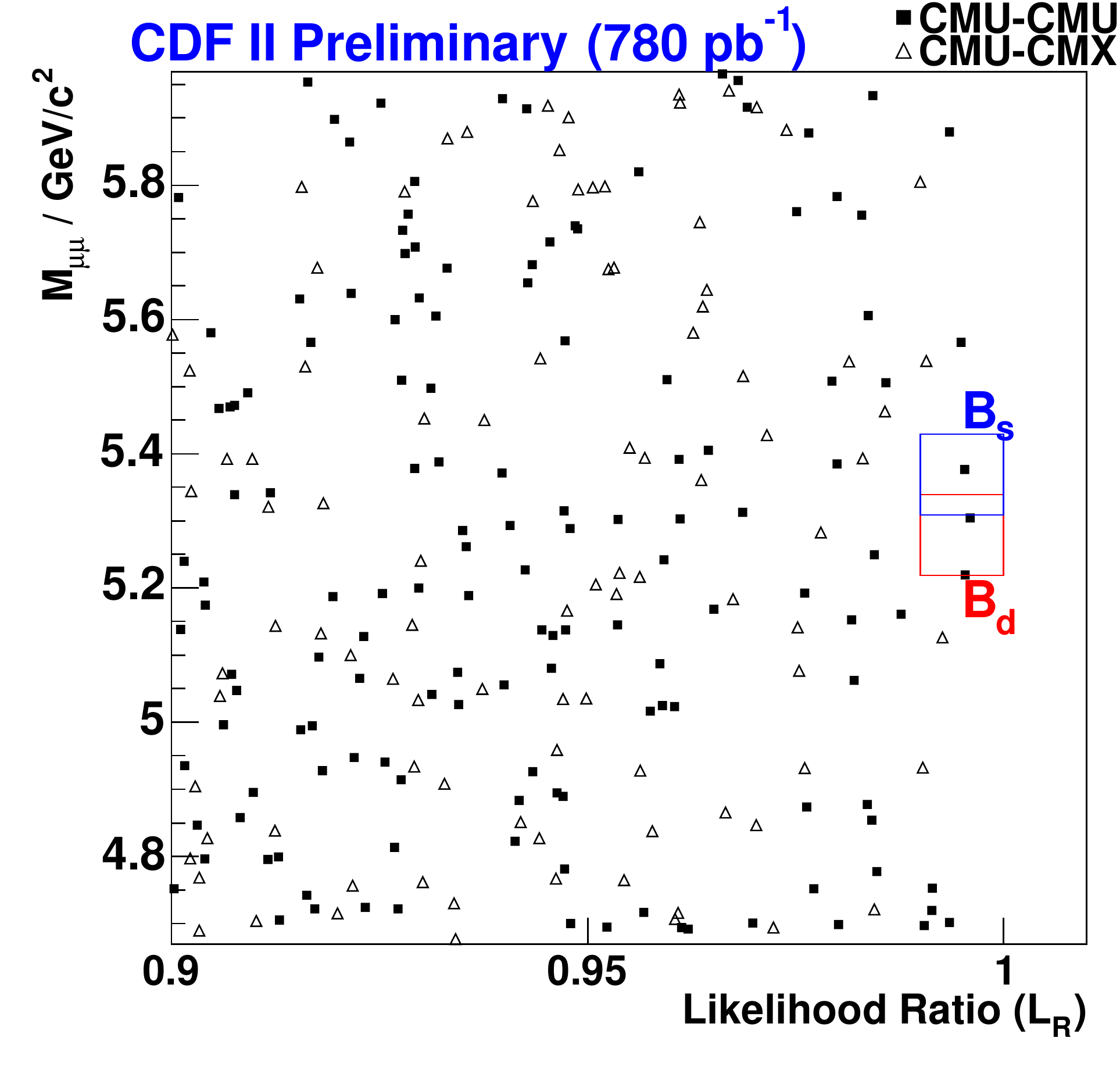}
\caption{\label{fig:bmmCDF}
The mass spectrum for \bpjpk\ candidate events (left),
and the distribution of likelihood value versus dimuon mass
for the \bmm\ candidate events (right).
}
\end{figure}

The \bpjpk\ events are required to
pass the same baseline selections, 
except for the dimuon mass to be consistent with \jp,
and the kaon track passing $p_T>1~\GeVc$.
After sideband subtraction and a small correction for \bpjpp, the number
of \bpjpk\ events is estimated to be $5763\pm 101$, as shown in Fig.~\ref{fig:bmmCDF}.

The following variables are used to discriminate signal from the background:
$I$ and $\Theta$ (defined above);
and the proper lifetime of the candidate $\lambda = c (\vec{L}\cdot\vec{p}) \Mmm/(\vec{p})^2$,
where $\vec{L}$ is the displacement of the decay vertex.
For a better discriminating power
variables $I$, $\Theta$, and $P(\lambda)\equiv \exp(-\lambda/c\tau_{B})$ 
(where $\tau_{B}$ is the world average \bsd\ lifetime)
are combined into a likelihood ratio  defined as
$L_R = \prod{P_s(x_i)}/(\prod{P_s(x_i)} + \prod{P_b(x_i)})$,
where $P_{s(b)}(x_i)$ is the probability for signal (background) of a variable $x_i$ 
(one of  the three).
The optimal cut is chosen to minimize the expected upper limit
on $\mathcal{B}(\bmm)$ at $90\%~\CL$,
and is found to be $L_R>0.99$.

Two background contributions are considered: 
combinatorial and mis-identification.
The mis-identification ($B\to h h^{(\prime)}$ reconstructed as \mm) is irreducible
and is estimated using the mis-identification rates (measured from data as in~\cite{Acosta:2003ag}) 
and assuming the same selection efficiency for $B\to h h^{(\prime)}$
and \bmm\ (except for the mass selection).

After applying the optimal cut $1.1\pm 0.4$ combinatorial background events 
are expected in the signal region, same
for \bs\ and \bd.
The mis-identification background is $0.2\pm 0.1$ events for \bs\ and $1.4\pm 0.2$ for \bd.
In data
1 (2) events pass the optimal cut in the \bs\ (\bd) signal window, as shown in Fig.~\ref{fig:bmmCDF},
corresponding to $\mathcal{B}(\bsmm)< 1.0\EE{-7}$ and $\mathcal{B}(\bdmm)< 3.0\EE{-8}$
at $95\%~\CL$
These results improve the previous results~\cite{Abulencia:2005pw,Bernhard:2005yn} 
by a factor of two and can be used to reduce the allowed parameter space of a broad spectrum
of SUSY models~\cite{Krutelyov:2005tc}.

Both CDF and D\O\ limits on \bsmm\ can be combined as independent measurements
to get a better limit~\cite{Bernhard:2005yn}.
With more data
it is expected that the combined sensitivity would reach the level of $1\EE{-8}$
by the end of Run~II~\cite{Krutelyov:2005tc}, which would make the \bsmm\ mode
a powerful probe in the search for the New Physics.

\section{Conclusion}
Study of rare decays of the heavy flavor is a substantial part of
the program of the \cdf\ and D\O\ experiments.
Owing to the high production cross section of all heavy flavor species
in the \ppbar\ collisions and to the efficient selection
methods it is possible to study rare
decays not available at other experiments as $B$-factories.
With the continuously increasing amount of data provided by the Tevatron
and improvements of the analyses the power of the experiments
continues to grow allowing for some of the world best results.
New results are available for FCNC decays in both charm and bottom
sectors with substantial improvements in the following upper limits:
\dpmmp\ ($\mathcal{B}< 4.7\EE{-6}$ at $90\%~\CL$), 
\bsmmphi\ ($\mathcal{B}< 3.2\EE{-6}$ at $90\%~\CL$), 
and \bmm\ ($\mathcal{B}(\bsmm)< 1 \EE{-7}$ and $\mathcal{B}(\bdmm)< 3 \EE{-8}$ 
at $90\%~\CL$) decays.
In addition, $D_s^+\to\phi\pi^+\to\mm\pi^+$ decay was observed with a significance
over $7\sigma$,
and evidence of the decay of $D^+$ into the same final state was reported with
a significance of $3.1\sigma$ and $\mathcal{B}(D^+\to\phi\pi^+\to\mm\pi^+)=(1.75\pm0.7\pm0.5)\EE{-6}$.
Improving sensitivity to \bsmm\ makes this decay one of
the most powerful probes of SUSY with large $\tan{\beta}$.
These results provide new insight into the properties of the FCNC decays,
which allows for improved tests of the SM and could
ultimately guide us to the New Physics.


\begin{theacknowledgments}
I would like to thank the members of CDF and D\O\ Collaborations
and especially the authors of the respective analyses.
I would also like to thank the organizers of the conference
for the opportunity to present results of these analyses.
\end{theacknowledgments}



\bibliographystyle{aipproc}   

\bibliography{hcp06-rare-decays}

\IfFileExists{\jobname.bbl}{}
 {\typeout{}
  \typeout{******************************************}
  \typeout{** Please run "bibtex \jobname" to obtain}
  \typeout{** the bibliography and then re-run LaTeX}
  \typeout{** twice to fix the references!}
  \typeout{******************************************}
  \typeout{}
 }

\end{document}

%% file: definitions.tex
\def\mrm{\mathrm}
\def\ra{\rightarrow}

\newcommand{\TeV}{\ensuremath{\mathrm{Te\kern -0.1em V}}}
\newcommand{\GeV}{\ensuremath{\mathrm{Ge\kern -0.1em V}}}
\newcommand{\MeV}{\ensuremath{\mathrm{Me\kern -0.1em V}}}
\newcommand{\keV}{\ensuremath{\mathrm{ke\kern -0.1em V}}}
\newcommand{\GeVc}{\ensuremath{\GeV/c}}
\newcommand{\GeVcc}{\ensuremath{\GeV/{c^2}}}
\newcommand{\MeVcc}{\ensuremath{\MeV/{c^2}}}

\newcommand{\cm}{\ensuremath{\mathrm{cm}}}
\newcommand{\um}{\ensuremath{\mathrm{\mu m}}}

\newcommand{\pbin}{\ensuremath{\mathrm{pb}^{-1}}}
\newcommand{\fbin}{\ensuremath{\mathrm{fb}^{-1}}}

\newcommand{\ase}[2]{\ensuremath{^{~+ #1}_{~- #2}}}

\newcommand{\bs}{\ensuremath{B_s^0}}
\newcommand{\bd}{\ensuremath{B_d^0}}
\newcommand{\bp}{\ensuremath{B^+}}

\newcommand{\kp}{\ensuremath{K^+}}
\newcommand{\jp}{\ensuremath{J/\psi}}
\newcommand{\mm}{\ensuremath{\mu^{+}\mu^{-}}}
\newcommand{\lplm}{\ensuremath{l^{+}l^{-}}}

\newcommand{\ee}{\ensuremath{e^{+}e^{-}}}
\newcommand{\pipi}{\ensuremath{\pi^{+}\pi^{-}}}

\newcommand{\bsd}{\ensuremath{B_{s(d)}^0}}
\newcommand{\bmm}{\ensuremath{B_{s(d)}^0\ra\mm}}
\newcommand{\bsmmphi}{\ensuremath{B_{s}^0\ra\phi\mm}}
\newcommand{\dmm}{\ensuremath{D^0\ra\mm}}
\newcommand{\dpp}{\ensuremath{D^0\ra\pipi}}
\newcommand{\dpmmp}{\ensuremath{D^+\ra\pi^+\mm}}
\newcommand{\dsmmp}{\ensuremath{D_{s}^+\ra\pi^+\mm}}
\newcommand{\dpmmk}{\ensuremath{D^+\ra K^+\mm}}
\newcommand{\dsmmk}{\ensuremath{D_{s}^+\ra K^+\mm}}

\newcommand{\bsmm}{\ensuremath{\bs\ra\mm}}
\newcommand{\bdmm}{\ensuremath{\bd\ra\mm}}
\newcommand{\bpjpk}{\ensuremath{\bp\ra\jp\kp}}
\newcommand{\bpjpp}{\ensuremath{\bp\ra\jp\pi^{+}}}

\newcommand{\EE}[1]{\ensuremath{\times 10^{#1}}}

\newcommand{\ppbar}{\ensuremath{p\overline{p}}}

\newcommand{\cdf}{CDF~II}

\newcommand{\CL}{\text{C.L.}}
\newcommand{\nbgd}{\ensuremath{N_{\mathrm{bgd}}}}

\newcommand{\Mmm}{\ensuremath{M_{\mu\mu}}}